\documentclass[aps,prb,twocolumn,floatfix,superscriptaddress,longbibliography]{revtex4-1}
\usepackage{graphicx,txfonts,bm}
\usepackage[colorlinks,citecolor=magenta,linkcolor=blue]{hyperref}
\usepackage{hyperref}

\begin{document}

\title{Tracing quasiparticle dynamics and hybridization dynamics in PuCoGa$_{5}$}

\author{Li Huang}
\email{lihuang.dmft@gmail.com}
\affiliation{Science and Technology on Surface Physics and Chemistry Laboratory, P.O. Box 9-35, Jiangyou 621908, China}

\author{Haiyan Lu}
\email{hyluphys@163.com}
\affiliation{Science and Technology on Surface Physics and Chemistry Laboratory, P.O. Box 9-35, Jiangyou 621908, China}
\date{\today}

\begin{abstract}
PuCoGa$_{5}$ has attracted significant attention due to its record-breaking superconducting transition temperature $T_c$=18.5~K among known $f$-electron superconductors. Here we systematically investigated the evolution of correlated electronic states in the plutonium-based unconventional superconductor PuCoGa$_{5}$ upon temperature using the embedded dynamical mean-field theory merged with density functional theory. The mixed-valence nature of PuCoGa$_{5}$ leads to intriguing quasiparticle dynamics and hybridization dynamics. Our findings reveal the presence of Dirac fermions and a temperature-driven localized-itinerant crossover of 5$f$ states. As the temperature decreases, the low-energy quasiparticle resonances develop gradually, while the high-energy quasiparticle resonances exhibit quite different behaviors with an initial increase and subsequent decrease. Furthermore, we identified a characteristic temperature of approximately 290~K for the onset of hybridization gaps, which is much lower than the coherence temperature 580~K for 5$f$ electrons. These results provide valuable insight on the the electronic structures, quasiparticle dynamics, and hybridization processes in $5f$ correlated electron systems.
\end{abstract}

\maketitle
\section{introduction}
The distinctive electronic structures of strongly correlated 4$f$-electron systems have attracted increasing interests in recent years. Among the most prominent examples are cerium-based heavy fermion materials, where the 4$f$ electrons are instrumental in shaping their fascinating lattice properties, such as quantum criticality, quantum magnetism, and heavy fermion superconductivity~\cite{RevModPhys.92.011002,CeCoIn5NatComm2023,PhysRevLett.100.087001,PhysRevLett.127.067002,PhysRevLett.126.216406,PhysRevLett.120.066403}. It is widely accepted that the 4$f$ electrons in these materials exhibit itinerant and localized dual nature. At low temperatures, the 4$f$ electrons become itinerant and are prone to hybridizing with conduction electrons through the Kondo screening mechanism, giving rise to heavy fermion states that lack magnetic order. In contrast, at high temperatures, the 4$f$ electrons tend to be incoherent, resulting in local moments and magnetism through the Ruderman-Kittel-Kasuya-Yosida (RKKY) mechanism. Consequently, cerium-based heavy fermion materials, akin to numerous other strongly correlated 4$f$-electron systems, are believed to undergo a crossover from high-temperature localized states to low-temperature itinerant states. However, the general scenario does not fully elucidate the development of collective hybridization with respect to temperature and its relationship with the emergence of heavy electrons. Recent experimental studies employing diverse probes have yielded conflicting results~\cite{PhysRevLett.124.057404,PhysRevB.96.045107}, underscoring the ongoing mystery and debate surrounding these complex systems.

In contrast to the typically localized 4$f$-electron systems characterized by large specific heat coefficients, the scenario for the strongly correlated $5f$-electron systems is notably more intricate~\cite{bauerreview}. Firstly, the temperature-driven 5$f$ localized-itinerant transitions in these systems have been scarcely explored, leaving it uncertain whether these transitions adhere to the same physical principles observed in 4$f$-electron systems. Secondly, it is commonly recognized that fluctuations in valence states are crucial for comprehending electronic correlation phenomena in 5$f$-electron systems. These fluctuations can result in atomic-multiplet-like quasiparticle resonances near the Fermi level. However, our understanding of how these quasiparticle resonances evolve with temperature remains limited. Thirdly, these quasiparticle resonances may interact with conduction electrons, generating multiple hybridization gaps. The dynamics of hybridization, including the orders and characteristic temperatures for the formation of these gaps, have yet to be thoroughly investigated.

PuCoGa$_{5}$ has garnered substantial attentions in the last few years due to its record-breaking superconducting transition temperature $T_c$ ($\approx 18.5$~K)~\cite{PhysRevLett.105.217002} among known $f$-electron superconductors. PuCoGa$_{5}$ crystallizes in a HoCoGa$_{5}$-type tetragonal structure as the cerium-based superconductors Ce$M$In$_5$ ($M$=Co, Rh, Ir) with space group $P4/mmm$~\cite{PhysRevLett.93.147005}. Its crystal structure consists of alternating PuGa$_{3}$ and CoGa$_{2}$ layers stacked along the $c$ axis [see Fig.~\ref{fig:akw}(a)]. Notably, two inequivalent Ga sites exist in the unit cell, labelled as Ga$_{1}$ and Ga$_{2}$, respectively. Ga$_{1}$ is located at the center of the $ab$ plane, while Ga$_{2}$ resides on the face of the unit cell. Both Ce$M$In$_5$ ($M$=Co, Rh, Ir)~\cite{PhysRevLett.84.4986,PhysRevLett.86.5152,PhysRevLett.86.4664,PhysRevLett.87.057002} and PuCoGa$_5$ are unconventional superconductors with the nodal $d$-wave symmetry of superconducting gap. A higher $T_c$ of PuCoGa$_5$ might be attributed to more itinerant 5$f$ electrons~\cite{PhysRevB.73.060506}, which is evidenced in weak magnetism measured by neutron scattering, X-ray magnetic circular dichroism and muon spin rotation experiments~\cite{PhysRevLett.100.076403,PhysRevLett.119.157204,HIESS20093210,HIESS2007709,OHISHI2007566}.
Furthermore, the experimental photoemssion spectra of PuCoGa$_{5}$~\cite{PhysRevLett.91.176401} exhibits a narrow peak near the Fermi level and a broad peak centered at -1.2 eV, primarily contributed mainly by 5$f$ electrons as well as hybridization between Pu-5$f$ and Co-3$d$ states, respectively. These narrow resonances are commonly observed in the Pu-based compounds, indicating partially itinerant nature of 5$f$ states.

The electronic structures of PuCoGa$_{5}$ has been investigated using the local spin density approximation (LSDA)~\cite{PhysRevLett.90.157001,PhysRevLett.90.207007,PhysRevB.70.104504,PIEKARZ20061029} and all electron full-potential linear muffin-tin orbitals method (FPLMTO)~\cite{PhysRevB.70.094515}. The experimental photoemssion spectra is roughly reproduced with the mixed level model which treats 5$f$ states into localized and itinerant parts~\cite{PhysRevLett.91.176401}. Even so, the itinerant and localized dual nature of 5$f$ electrons can not be accurately described within traditional density functional theory. The prominent peak in the experimental photoemssion spectra can be accurately reproduced using density functional theory combined with dynamical mean-field theory (DFT+DMFT) which takes 5$f$ correlations into account~\cite{PhysRevB.73.060506,PhysRevB.98.035143}. Numerous experimental and theoretical investigations have been conducted to study the $5f$ electronic structures and pairing mechanism of PuCoGa$_{5}$~\cite{PhysRevB.72.014521,UMMARINO20094}. Previous works suggested that the 5$f$ electrons in PuCoGa$_{5}$ reside in the itinerant side, in contrast to the other isostructural plutonium-based ``115'' materials, such as PuCoIn$_{5}$~\cite{Zhu2012EPL,Bauer2011JPCM}. This makes PuCo(Ga, In)$_{5}$ a particularly suitable platform for studying the 5$f$ localized-itinerant transition driven by alloying. However, the impact of temperature on its $5f$ electronic structure remains less explored.

In the present paper, we utilize the density functional theory in combination with the embedded dynamical mean-field theory~\cite{RevModPhys.68.13,RevModPhys.78.865,PhysRevB.81.195107} to investigate the temperature-dependent evolution of electronic structures in PuCoGa$_{5}$. We not only identify a Dirac cone in its band structure, but also observe a 5$f$ localized-itinerant crossover. Furthermore, our findings indicate that the valence state fluctuations in PuCoGa$_{5}$ could significantly influence its quasiparticle dynamics and hybridization processes, leading to distinct quasiparticle resonances and multi-stage hybridizations.

\section{Methods}
A combination of the density functional theory and the embedded dynamical mean-field theory (DFT + DMFT) is employed to treat the many-body interaction between 5$f$ electrons of PuCoGa$_5$. The DFT + DMFT method integrates the realistic band structure calculation from DFT and a non-perturbative way to tackle the many-body local interaction effects in DMFT~\cite{RevModPhys.68.13,RevModPhys.78.865}. Given that the relativistic effect is crucial to the electronic structure for Pu atom and the spin-orbit coupling is approximately 430 meV~\cite{shim:2007}, which exceeds the crystal field splitting, we have chosen to neglect the crystal field splitting in our calculations.

\textbf{DFT calculations.} We employed the \texttt{WIEN2K} code to perform the DFT calculations, which implements a full-potential linearized augmented plane-wave formalism~\cite{wien2k}. The experimental crystal structure of PuCoGa$_5$ was adopted~\cite{PhysRevLett.93.147005}, ignoring the thermal expansion effects within the temperature range. The muffin-tin sphere radii for Pu, Co and Ga atoms were chosen 2.50 au, 2.38 au and 2.26 au, respectively, and $R_{\text{MT}}K_{\text{MAX}} = 8.0$. The valence states included the $5f$, $6d$, and $7s$ orbitals for Pu, the $3d$ and $4s$ orbitals for Co, and $4f$, $5d$, and $6s$ orbitals for Ga. The exchange-correlation potential was evaluated using the Generalized Gradient Approximation (GGA) with the Perdew-Burke-Ernzerhof (PBE) functional~\cite{PhysRevLett.77.3865}. The spin-orbit coupling effect was included in a variational manner. The PuCoGa$_5$ was treated as nonmagnetic according to the results of neutron scattering experiment~\cite{HIESS2007709}.

\textbf{DFT + DMFT calculations.} We utilized the \texttt{eDMFT} software package to perform the DFT + DMFT calculations~\cite{PhysRevB.81.195107}. The many-body nature of the Pu-$5f$ orbitals was captured by the DMFT formalism. The Coulomb interaction matrix for Pu-$5f$ orbitals was constructed by using the Slater integrals. The Coulomb repulsion interaction parameter $U$ and Hund's exchange interaction parameter $J_{\text{H}}$ were 5.0~eV and 0.6~eV~\cite{PhysRevB.101.125123}, respectively. We used the $|J, J_z\rangle$ basis to construct the local impurity Hamiltonian. A large energy window, from -10~eV to 10~eV with respect to the Fermi level, was used to build the DMFT projector, which was used to project the Kohn-Sham basis to local basis. The vertex-corrected one-crossing approximation (OCA) impurity solver~\cite{PhysRevB.64.155111} was employed to solve the resulting multi-orbital Anderson impurity models. In order to reduce the computational consumes and accelerate the calculations, we truncated the Hilbert space of the local impurity problems, retaining only contributions from those atomic eigenstates with $N \in [3,7]$. Then all results concerning spectral functions and self-energies were obtained using the OCA impurity solver. A double-counting term was included to account for the excess electronic correlations that were already partially incorporated in the DFT calculations. Here we chose the fully localized limit scheme to describe the double-counting term~\cite{jpcm:1997}, which is given by 
\begin{equation}
\Sigma_{\text{dc}} = U \left(n_{5f} - \frac{1}{2}\right) - \frac{ J_{\text{H}} } {2} \left(n_{5f} -1 \right),
\end{equation}
where $n_{5f}$ is the nominal occupancy of Pu-$5f$ orbitals, fixed at 5.2 during the DFT + DMFT calculations. A charge fully self-consistent DFT + DMFT calculations are performed, requiring approximately $60 \sim 80$ DFT + DMFT iterations to achieve good convergence. The convergence criteria for charge density and total energy were set to $10^{-5}$~e and $10^{-5}$~Ry, respectively. Notably, the direct output of OCA impurity solver is real axis self-energy $\Sigma (\omega)$, which was used to calculate the momentum-resolved spectral functions $A(\mathbf{k},\omega)$, density of states $A(\omega)$, renormalization factor $Z$, effective electron mass $m^{\star}$, Fermi surface and other physical observables. For the three-dimensional Fermi surface, the energy bands that cross the Fermi level were plotted as isoenergetic surface. Additionally, the two-dimensional Fermi surface was depicted only on the $k_x-k_y$ plane with $k_z = \pi/2$, providing insight into the internal geometry of the three-dimensional Fermi surface.

\section{Results}
\subsection{Photoemssion spectra} 
\begin{figure}[ht]
\includegraphics[width=0.45\textwidth]{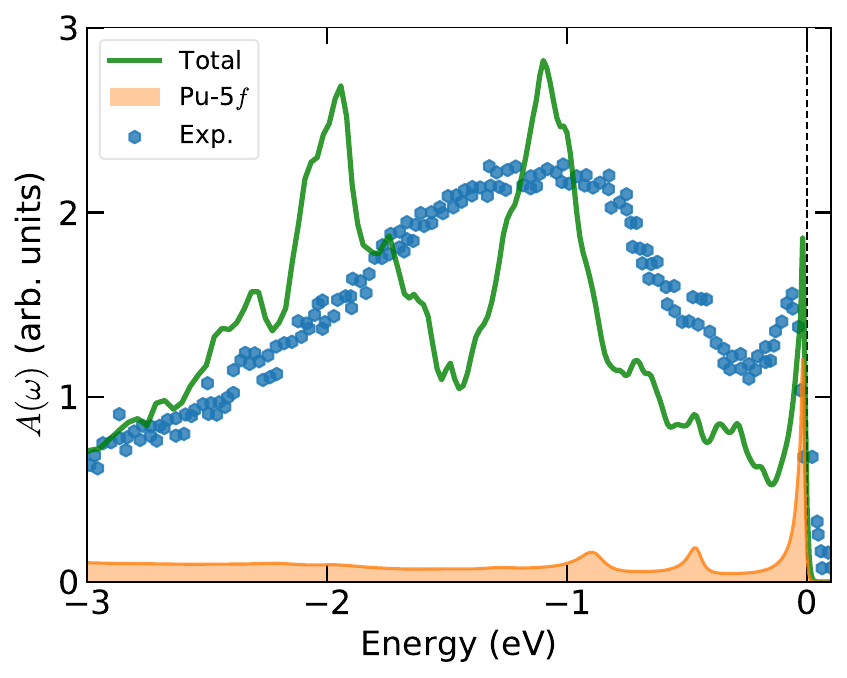}
\caption{(Color online). The total density of states (green-solid line) and 5$f$ partial density of states (color-filled regions) of PuCoGa$_{5}$ obtained by the DFT + DMFT method, as well as replotted experimental photoemssion spectra at 77~K~\cite{PhysRevLett.91.176401}. \label{fig:comp}}
\end{figure}

In order to validate reliablity of our calculations, we compared our results with experimental photoemssion spectra of PuCoGa$_{5}$~\cite{PhysRevLett.91.176401} [see Fig.~\ref{fig:comp}]. Since the experimental data were collected at 77~K, the calculated total density of states was multiplied by a Fermi-Dirac distribution function $f(\epsilon)=1/[exp(\epsilon\beta)+1]$ with inverse temperature $\beta$=150 ($T\approx77$~K). Evidently, the calculated total density of states captures the salient features of experimental photoemssion spectra, encompassing a narrow peak proximate to the Fermi level and a broad peak centered at -1.2 eV, primarily contributed mainly by 5$f$ electrons as well as hybridization between Pu-5$f$ and Co-3$d$ states, respectively. Furthermore, the characteristic peaks in the total density of states align well with previous theoretical studies~\cite{PhysRevB.73.060506}, thereby affirming the credibility of our calculations.

\subsection{Dirac fermions} 
\begin{figure*}[ht]
\includegraphics[width=\textwidth]{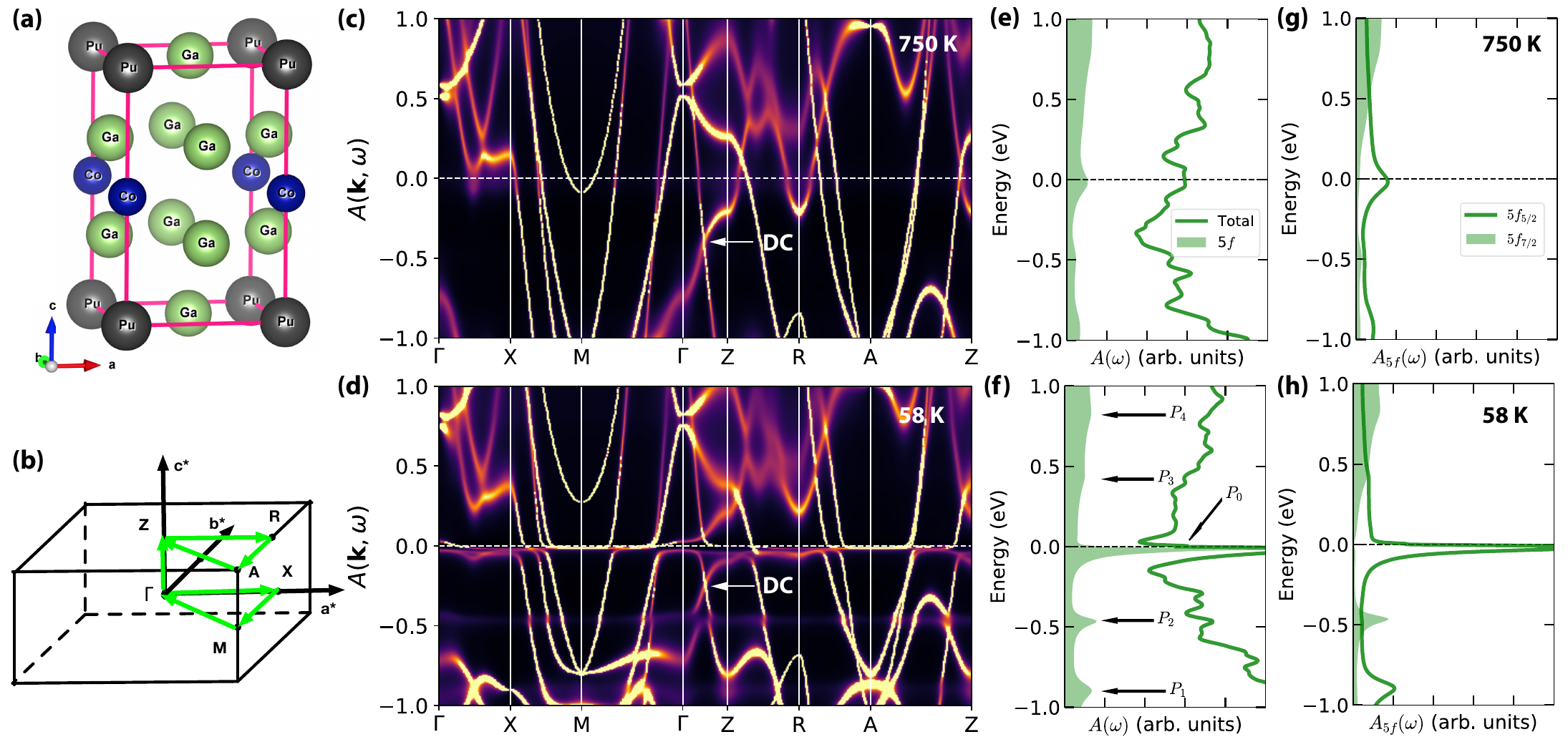}
\caption{(Color online). Temperature-dependent electronic structures of PuCoGa$_{5}$. (a) Crystal structure of PuCoGa$_{5}$. Note that the Pu, Co, and Ga atoms are represented by grey, blue, and green spheres, respectively. (b) Schematic picture for the first Brillouin zone of PuCoGa$_{5}$. The green arrows are used to visualize the selected high-symmetry directions. (c)-(d) Quasiparticle band structures at 750~K and 58~K. Along the $\Gamma-Z$ line, the Ga$_{2}$-$p_z$ and Ga$_{1}$-$s$ orbitals cross with each other, which leads to a Dirac cone, as indicated by white arrow. (e)-(f) Total density of states $A(\omega)$ and partial 5$f$ density of states $A_{5f}(\omega)$ at 750~K and 58~K. Here, the labels $P_{0}$, $P_{1}$, $P_{2}$, $P_{3}$, and $P_{4}$ denote the peaks of quasiparticle multiplets. (g)-(h) Partial 5$f$ density of states split by spin-orbit interactions at 750~K and 58~K. \label{fig:akw}}
\end{figure*}

The fingerprints of Dirac fermions have been identified in the well-known lanthanum- and cerium-based ``115'' systems, specifically in compounds of the form $RM$In$_{5}$, where $R=$ La and Ce, $M=$ Co, Rh, and Ir~\cite{Guo2021,kent:2018}. These compounds share the same crystal symmetry with PuCoGa$_{5}$. This raises a pertinent question. Can Dirac fermions be detected in PuCoGa$_{5}$? Figure~\ref{fig:akw}(c)-(d) illustrate the momentum-resolved spectral functions $A(\mathbf{k},\omega)$ of PuCoGa$_{5}$ for two representative temperatures along some selected high-symmetry directions [see Fig.~\ref{fig:akw}(b)]. Regardless of temperature, there are indeed lots of bands that crossing the Fermi level. They likely belong to the $spd$ conduction electrons and exhibit significant dispersions. Especially, along the $\Gamma-Z$ line, an accidental band crossing occurs approximately 300 $\sim$ 400~meV below the Fermi level, resulting in a symmetry protected Dirac node. Through detailed band analysis, we discern that the two Dirac bands predominantly consist of Ga$_{2}$-$p_{z}$ (band representation $\Lambda_7$) and Ga$_{1}$-$s$ (band representation $\Lambda_6$), respectively. Remarkably, this Dirac point maintains its integrity across a range of temperatures. However, a discernible shift occurs in its energy position, systematically moving closer to the Fermi level as the temperature is lowered.

\begin{figure*}[ht]
\includegraphics[width=0.8\textwidth]{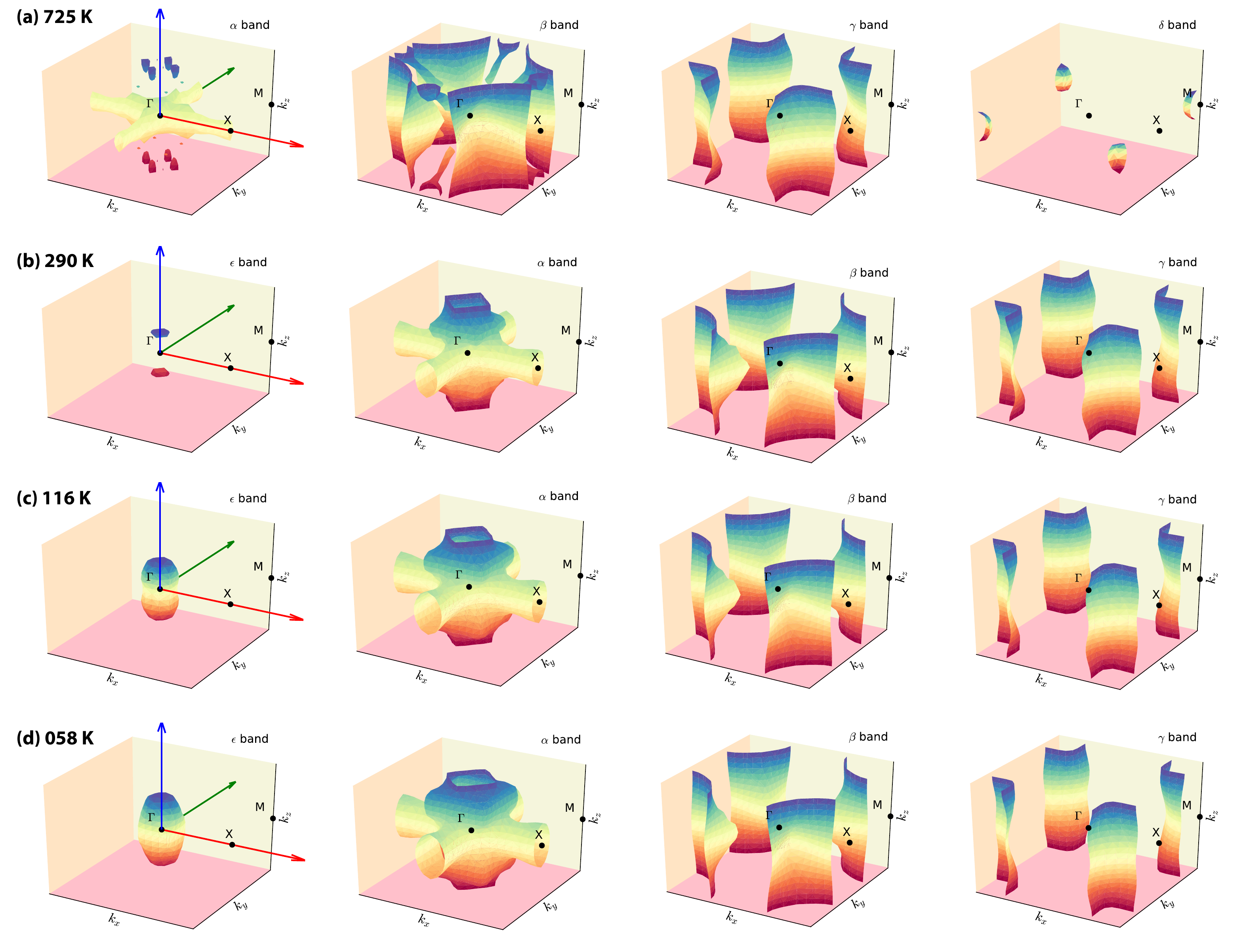}
\caption{(Color online) 3D Fermi surfaces of PuCoGa$_5$ obtained by the DFT + DMFT method. (a) $T = 725$~K. (b) $T = 290$~K. (c) $T = 116$~K. (c) $T = 58$~K. There are four doubly degenerated bands crossing the Fermi level. \label{fig:fs3d}}
\end{figure*}

\begin{figure*}[ht]
\includegraphics[width=\textwidth]{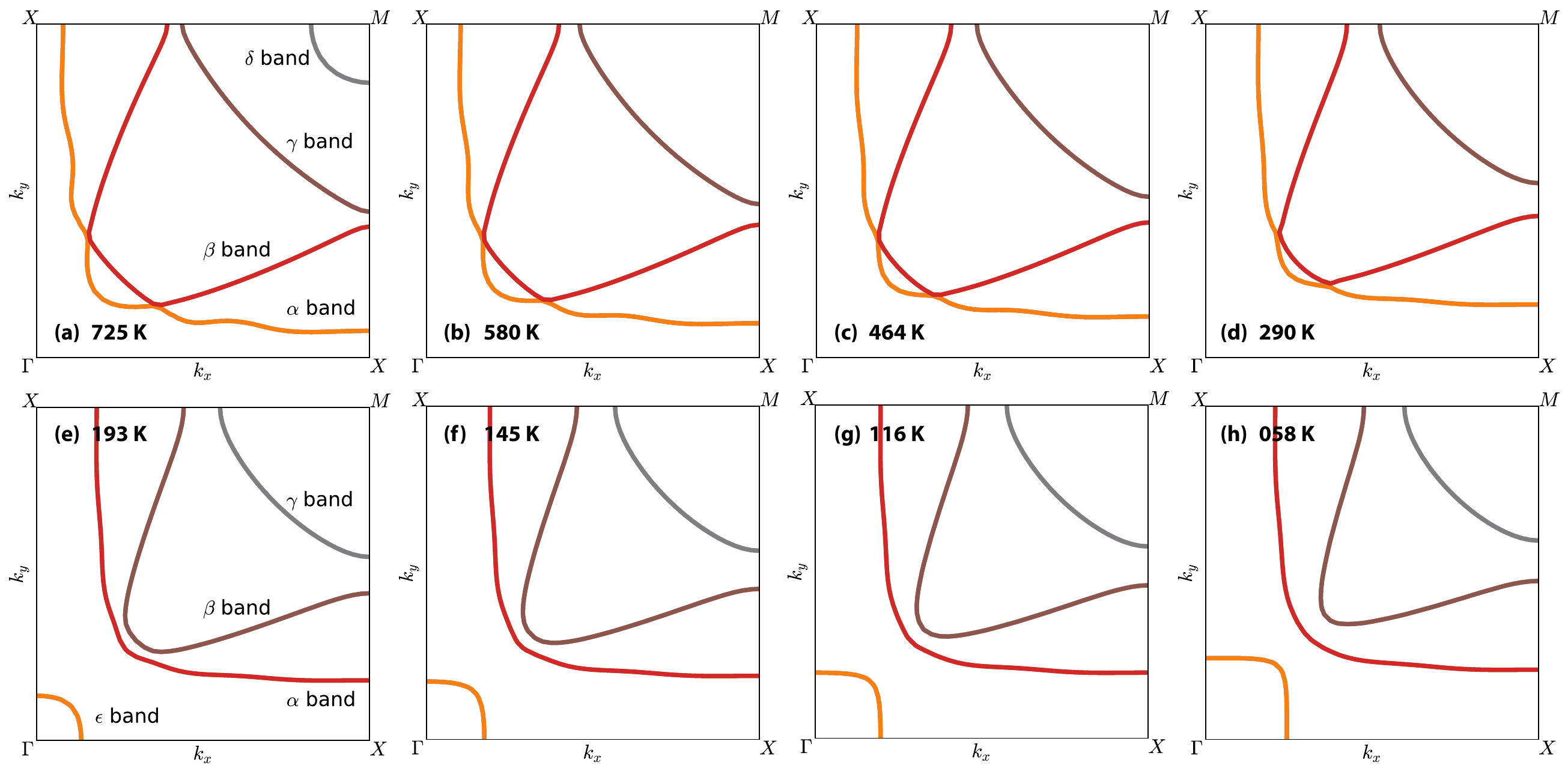}
\caption{(Color online) 2D Fermi surfaces of PuCoGa$_5$ calculated by the DFT + DMFT method. Only the $k_x-k_y$ plane (with $k_z = \pi/2$) which intersects the $\Gamma$-$X$-$M$ cross section in Fig.~\ref{fig:fs3d} is shown. There are four doubly degenerated bands traversing the Fermi level. They are visualized using different colors. \label{fig:fs2d}}
\end{figure*}

\subsection{Coherence temperature of 5\texorpdfstring{$f$}{f} electrons} 

In comparision with Ga's $sp$ electrons, the behavior of Pu's $5f$ electrons is far more susceptible to temperature variations [see Fig.~\ref{fig:akw}(c)-(d)]. At high temperature ($T = 750$~K), the $c-f$ hybridization is negligible, resulting in a band structure for quasiparticles that appears largely incoherent. Conversely, at lower temperature ($T = 58$~K), the 5$f$ electrons form nearly flat bands below the Fermi level. They strongly hybridized with the conduction electrons, rendering the heavily distorted band structures near the Fermi level. These changes are also visible in the electronic density of states. In Figure~\ref{fig:akw}(e)-(f) illustrates the total density of states $A(\omega)$ and partial 5$f$ density of states $A_{5f}(\omega)$ of PuCoGa$_{5}$. At high temperature, the spectrum $A_{5f}(\omega)$ presents only a minor ``hump'' near the Fermi level, indicative of the localized nature of the $5f$ electrons. Nevertheless, at low temperature, this ``hump'' evolves into a distinct Kondo resonance peak, signifying the transition of the 5$f$ electrons from localized to itinerant behavior and the emergence of heavy electron states. In essence, a temperature-driven 5$f$ localized-itinerant crossover would occurs in PuCoGa$_{5}$, concurrent with a transition from a small to a large Fermi surface (i.e. an electronic Lifshitz transition).

Figure~\ref{fig:fs3d} displays the temperature-dependent three-dimensional Fermi surface structure which can be detected through quantum oscillation and angle-resolved photoemission spectroscopy experiments. Initially, at high temperature of 725~K, the geometries of four Fermi surfaces ($\alpha$, $\beta$, $\gamma$, $\delta$) are evidently different [see Fig.~\ref{fig:fs3d}(a)]. The $\alpha$ band resembles fragmentation, the $\beta$ band is akin to four cylindrical sections accompanied by four small slices, the $\gamma$ band consists of four cylindrical electron sheet centered at the $M$ point, and the $\delta$ band exhibits minimal distribution confined to the four corners. Particularly, the 5$f$ states are predominantly localized at high temperatures, with only a minor spectral weight near the Fermi level. Upon cooling to 290~K, PuCoGa$_5$ undergoes a significant transformation: the $\delta$ band vanishes, accompanied by the emergence of $\epsilon$ band [see Fig.~\ref{fig:fs3d}(b)]. The $\alpha$ band transforms from a fragmented Fermi surface structure at high temperatures to a deformed hexahedral structure. The $\beta$ band loses its four small slices, maintaining the four cylindrical sections. Meanwhile, the $\gamma$ band retains its shape and volume at high temperatures. It is worth mentioning that the $\epsilon$ band has two small pieces along the $k_z$ axis. As shown in Fig.~\ref{fig:fs3d}(c)-(d) at temperatures of 116~K and 58~K, the three Fermi surfaces ($\alpha$, $\beta$, $\gamma$) remain largely unchanged, while the $\epsilon$ band continuously expands in volume as the temperature decreases. This correspondence aligns well with previous density functional theory (DFT) calculations~\cite{PhysRevLett.90.207007}, featuring cylindrical Fermi sheets similar to those found in Ce$M$In$_5$ ($M$=Co, Rh, Ir) compounds. Consequently, the evolution of the three-dimensional Fermi surfaces with temperature substantiates a temperature-driven localized-delocalized crossover of 5$f$ states.

To further characterize the internal structure of the three-dimensional Fermi surface, Fig.~\ref{fig:fs2d} displays eight representative temperature points of two-dimensional Fermi surface structures. The electronic band structure at 725~K is distinct from the others, as the $\delta$ band located at the gray line point vanishes below this temperature. At temperatures of 580~K, 464~K, and 290~K, the band structures are essentially identical. Additionally, the band structures at 193~K, 145~K, 116~K, and 58~K aslo exhibit significant similarity. Below 290~K, an orange $\epsilon$ band emerges at the $\Gamma$ point, with its enclosed area gradually expanding as the temperature decreases. Thus, the evolution of two-dimensional and three-dimensional Fermi surfaces reveals a localized-delocalized crossover of 5$f$ states below a critical temperature of approximately 580~K. This crossover coincides with the appearance of the central quasiparticle resonance $P_0$, which exceeds the temperature of $c-f$ hybridization gap opening at 290~K, as well as the characteristic temperature of the quasiparticle resonance peak at 300~K.

\begin{figure*}[ht]
\includegraphics[width=1.0\textwidth]{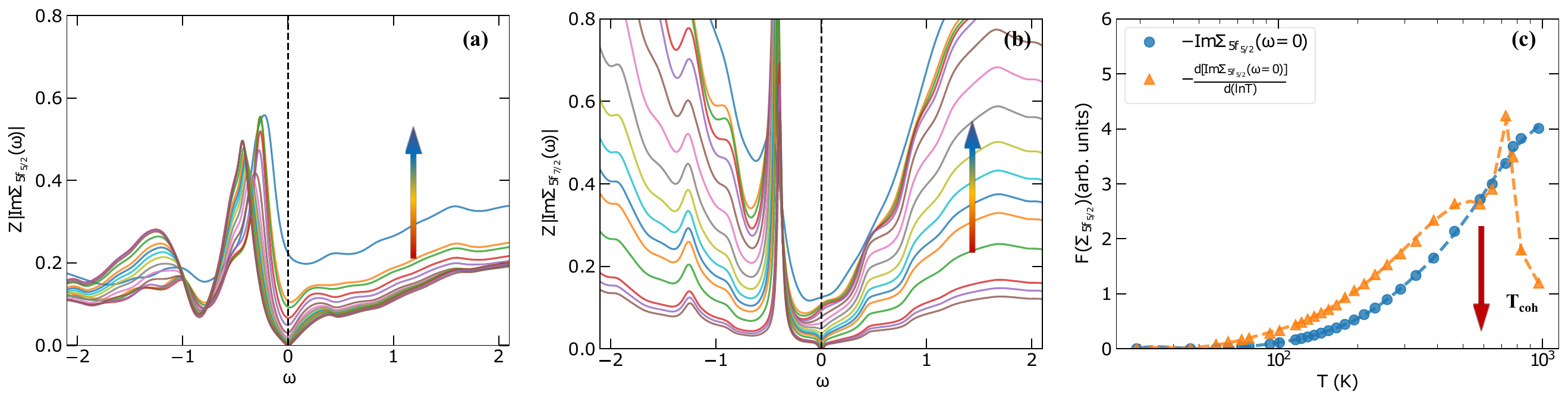}
\caption{(Color online). Real-frequency self-energy functions of PuCoGa$_5$ obtained by the DFT + DMFT method. (a)-(b) Temperature-dependent $Z|\text{Im}\Sigma(\omega)|$ for the $5f_{5/2}$ and $5f_{7/2}$ states. $Z$ means the renormalization factor. In these panels, the arrows denote the increasing system temperature ($T$ is from 29 K to 970 K). (c) $-d\text{Im}\Sigma_{5f_{5/2}}(\omega = 0) / d\text{ln}T$ and $-\text{Im}\Sigma_{5f_{5/2}}(\omega=0)$ as a function of temperature $T$, where the red arrows indicate the coherent temperatures [$T_{\text{coh}}(5f_{5/2}) \approx 580$~K].
\label{fig:sig}}
\end{figure*}

Due to the intricate spin-orbit coupling, the 5$f$ orbitals are divided into two manifolds, the $5f_{5/2}$ and $5f_{7/2}$ states. Their behaviors differ significantly with temperature variations. As illustrated in Fig.~\ref{fig:akw}(g)-(h), the $5f_{5/2}$ state demonstrates a clear temperature dependence, transitioning gradually from localized to itinerant as the temperature decreases. This transformation leads to a prominent Kondo resonance peak and the emergence of a substantial Fermi surface. The coherence temperature is also derived from the self-energy function which encodes the electron correlation effects~\cite{RevModPhys.78.865,RevModPhys.68.13}. $\Sigma(\omega)$ is utilized to determine the quasiparticle weight $Z$ and effective electron masses $m^{*}$ for Pu-$5f$ electrons~\cite{RevModPhys.68.13}:
\begin{equation}
\label{eq:renor}
Z^{-1} = \frac{m^{*}}{m_e} = 1 - \frac{\partial \text{Re}\Sigma(\omega)}{\partial \omega}\Big|_{\omega = 0},
\end{equation}
where $m_e$ denotes the mass of non-interacting electron. Figures~\ref{fig:sig}(a)-(b) show the renormalized imaginary parts of self-energy functions $Z|\text{Im}\Sigma(\omega)|$ for the $5f_{5/2}$ and $5f_{7/2}$ states, respectively. Here $Z$ represents the quasiparticle weight or renormalization factor, which signifies the strength of the electron correlation. Specifically, $Z|\text{Im}\Sigma(0)|$ reflects the low-energy electron scattering rate~\cite{PhysRevB.99.125113}, which decreases rapidly as the temperature drops. This results in longer quasiparticles lifetimes and onset of coherence, producing a decrease of electrical resistivity below coherence temperature~\cite{PhysRevLett.123.217002}. At low temperatures, both $Z|\text{Im}\Sigma_{5f_{5/2}}(0)|$ and $Z|\text{Im}\Sigma_{5f_{7/2}}(0)|$ approach zero. With an increase in temperature, they attain finite values and rise rapidly. Notably, the quantitative coherent temperature can be extracted from the self-energy functions.

Figure~\ref{fig:sig}(c) displays the imaginary parts of self-energy functions at the Fermi energy for the $5f_{5/2}$ states, specifically $-\text{Im}\Sigma_{5f_{5/2}}(\omega = 0)$, along with the first derivative with respect to temperature, -$d\text{Im}\Sigma_{5f_{5/2}}(\omega = 0) / d\text{ln}T$. In comparison with Ce-based compounds~\cite{Shim1615} and UTe$_2$~\cite{PhysRevLett.123.217002}, -$d\text{Im}\Sigma_{5f_{5/2}}(\omega = 0) / d\text{ln}T$ and its peak corresponds to the coherent temperature $T_{\text{coh}}$. When $T < T_{\text{coh}}$, the electron coherence emerges and heavy electron states develop quickly, leading to the rapid growth of the quasiparticle peak near the Fermi level~\cite{RevModPhys.92.011002}. It is deduced that $T_{\text{coh}}$ is about 580~K for the $5f_{5/2}$ state, which is roughly consistent with the value derived from the variation in Fermi surface geometry [Fig.~\ref{fig:fs2d} and Fig.~\ref{fig:fs3d}]. This suggests that the coherence temperature shall provide valuable insights for future electrical resistivity measurements. 

\subsection{Evolution of quasiparticle resonances} 
\begin{figure*}[ht]
\includegraphics[width=\textwidth]{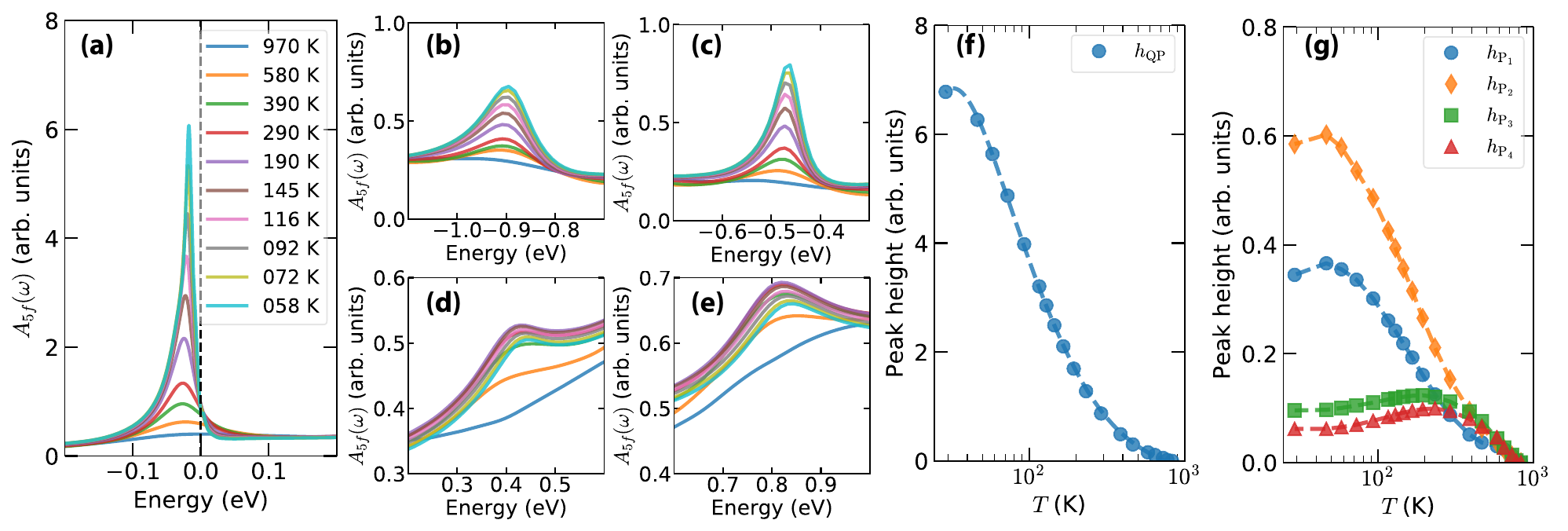}
\caption{(Color online). Quasiparticle dynamics in PuCoGa$_{5}$. (a)-(e) Evolutions of the quasiparticle multiplets ($P_{0}$, $P_{1}$, $P_{2}$, $P_{3}$, and $P_{4}$) with respect to temperature. (f) Height of the central quasiparticle peak ($P_0$) as a function of temperature $T$. (g) Heights of the quasiparticle multiplets peaks ($P_1$, $P_{2}$, $P_{3}$, and $P_{4}$) as a function of temperature $T$. In this panel, the heights at $T_{\text{max}}$ ($\approx 970$~K) are assumed to be zero, and they are used to calibrate the values at lower temperatures. \label{fig:dos}}
\end{figure*}

Many Pu-based correlated electrons materials, such as $\alpha$-Pu, $\beta$-Pu, and PuTe, are classified as Racah metals, exhibiting stripe-like features in quasiparticle band structures and multiple spur peaks in the density of states near the Fermi level. These signatures demonstrate the presence of quasiparticle multiplets, which arise from valence state fluctuations and many-body transitions among various atomic eigenstates. As is clearly seen in Fig.~\ref{fig:akw}(d), there are stripe-like bands at about 400~meV and 900~meV below the Fermi level, accompanied by corresponding peaks in the density of states shown in Fig.~\ref{fig:akw}(f). Consequently, at low temperatures, PuCoGa$_{5}$ is indeed a realization of the Racah metal. As shown in Fig.~\ref{fig:akw}(f), in addition to the central quasiparticle peak $P_{0}$, there are additional four satellite peaks (labelled as $P_{1}-P_{4}$). Since these peaks are strongly temperature-dependent, they are classified as quasiparticle resonances. 

The temperature dependence of these quasiparticle resonances are depicted in Fig.~\ref{fig:dos}. As for the central quasiparticle resonance $P_0$, it appears when $T < $ 580~K and grows monotonically with decreasing temperature [see Fig.~\ref{fig:dos}(a) and (f)]. However, the other quasiparticle resonances ($P_1 - P_4$) exhibit non-monotonic behavior. As the temperature decreases, they develop quickly at first and then decay slowly at moderate temperatures $T^{*}$. As is shown in Fig.~\ref{fig:dos}(g), at $T^{*} \approx 300$~K and $T^{*} \approx 50$~K, the heights of $P_1$ ($P_2$) and $P_{3}$ ($P_{4}$) reach their maximum values, respectively. The doublet of peaks $P_{3}$ and $P_{4}$ can be viewed as a reflection of $P_1$ and $P_2$ about the Fermi level. Their heights approach some saturated values when $T$ approaches zero. The evolution of low-energy quasiparticle resonance ($P_0$) differs from those of high-energy quasiparticle resonances ($P_{1}-P_{4}$).

The emergence-and-decay behaviors of quasiparticle resonances $P_{1} - P_{4}$ can be explained by using the mixed-valence picture. Similar to Pu and PuTe, PuCoGa$_{5}$ is a mixed-valence system with an average 5$f$ occupancy of $\langle n_{5f} \rangle \approx 5.09$. Strong valence state fluctuations between the $5f^{5}$ and $5f^{6}$ configurations contribute to these quasiparticle resonances. Their weights are related to the transition probabilities of the underlying atomic eigenstates, while their positions are determined by the energy spacings between the excited and the ground atomic eigenstates. Undoubtedly, the dominant electronic configuration in PuCoGa$_{5}$ is always $5f^{5}$. However, as the temperature decreases, the $5f$ electrons spend more and more their lifetimes in the $5f^{6}$ configuration, which enhances the $5f^{5}-5f^{6}$ many-body transitions, leading to the appearances of $P_1 - P_{4}$. Conversely, when $T < T^{*}$, the energy spacings are enlarged, and the transitions are suppressed, causing these quasiparticle resonances $P_1 - P_4$ to decay and their positions to shift.

\begin{figure*}[ht]
\includegraphics[width=\textwidth]{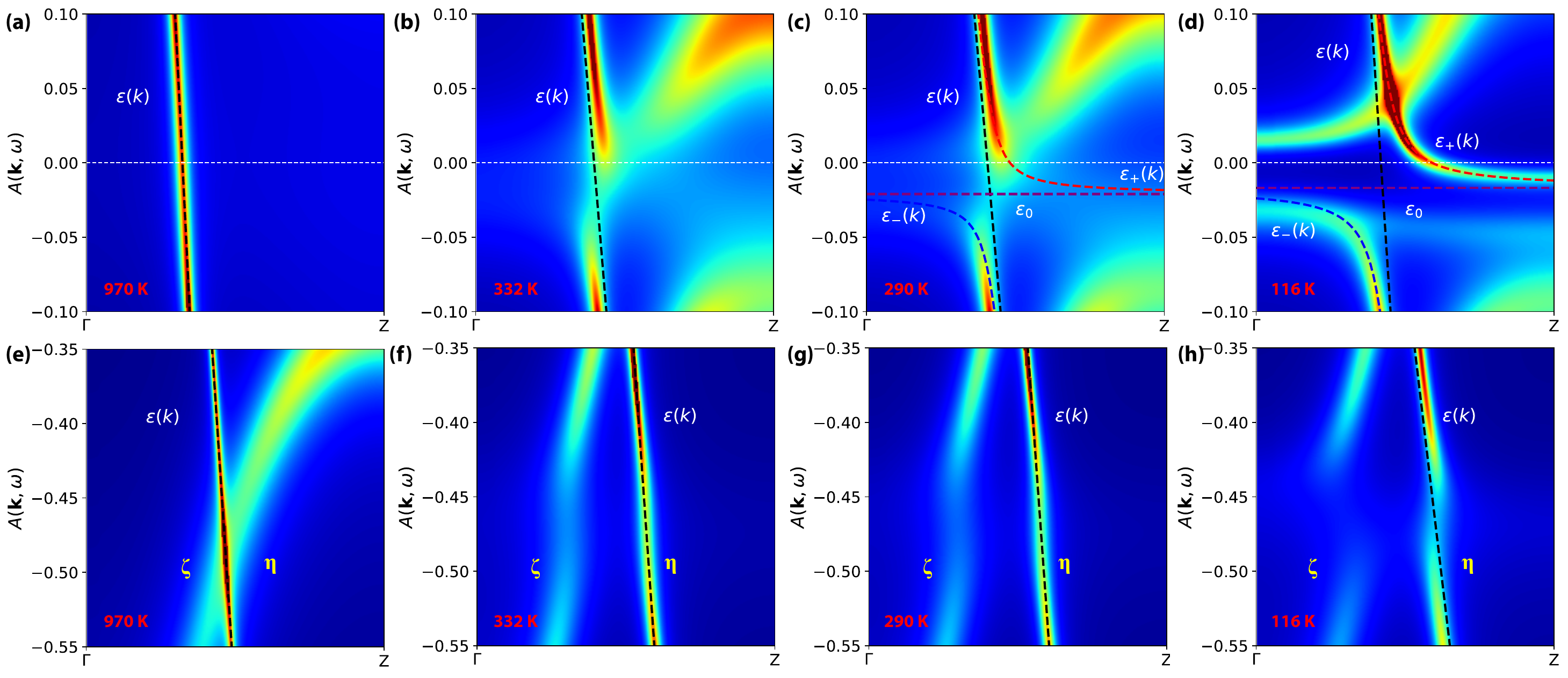}
\caption{(Color online). Quasiparticle band structures of PuCoGa$_{5}$ along the $\Gamma-Z$ line. The white horizontal dashed line means the Fermi level. The other colorful dashed lines denote the band fitting results by using the periodical Anderson model. See main text for more details. \label{fig:hyb}}
\end{figure*}

\begin{figure}[ht]
\includegraphics[width=\columnwidth]{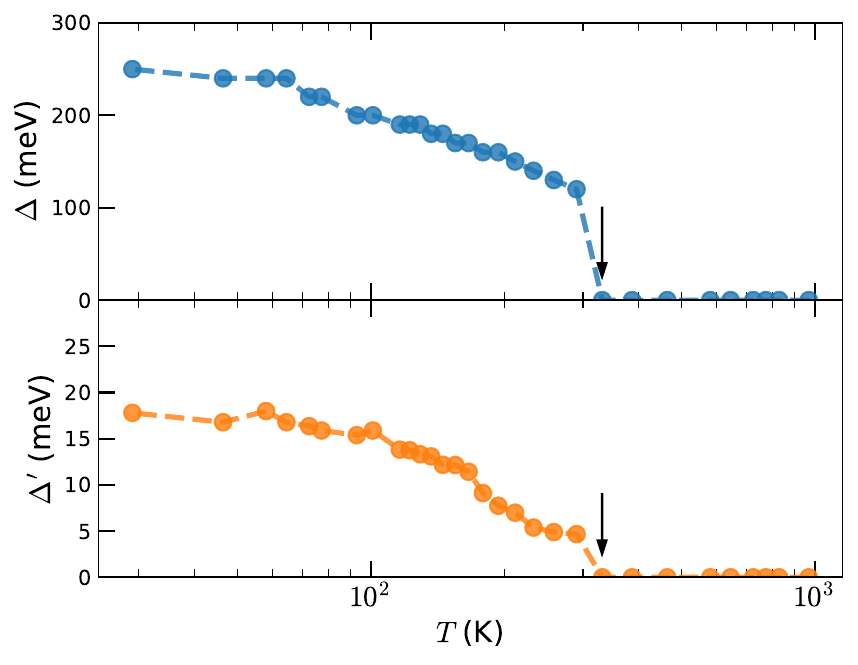}
\caption{(Color online). The temperature-dependent direct and indirect hybridization gaps ($\Delta$ and $\Delta'$) of PuCoGa$_{5}$ along the $\Gamma-Z$ line. The vertical arrows are used to indicate the temperatures in which the gaps are opened. \label{fig:gap}}
\end{figure}

\subsection{Hybridization dynamics}
In PuCoGa$_{5}$, strong valence state fluctuations not only give rise to intricate quasiparticle dynamics, but also result in complex hybridization dynamics. As mentioned before, there are at least five quasiparticle resonances ($P_0 - P_{4}$). Each of them is linked to a distinct hybridization process. It is anticipated that these resonances possess unique characteristic energy scales, such as the onset temperatures for hybridization and the opening of hybridization gaps. In Fig.~\ref{fig:hyb}, the quasiparticle band structures in the $\Gamma - Z$ directions at various temperatures are visualized, with $P_0$ and $P_2$ serving as illustrative cases for the hybridization dynamics.

For $P_0$ [see Fig.~\ref{fig:hyb}(a)-(d)], considerable band distortions are observed at $T = 332$~K, indicating the onset of $c-f$ hybridization. At room temperature, the hybridization gap is already present, which expands as the temperature declines. To estimate the hybridization gap and monitor its temperature dependence, a simple mean-field model, namely, the periodical Anderson model~\ref{eq:pam}, is employed to fit the low-energy hybridized bands.

\begin{equation}
\label{eq:pam}
E_{\pm}(k) = \frac{[\epsilon_0 + \epsilon(k)] \pm \sqrt{[\epsilon_0 - \epsilon(k)]^2 + 4|V_k|^2}}{2},
\end{equation}
where $\pm$, $\epsilon_0$, $\epsilon(k)$, and $V_k$ denote the upper and lower band branches, renormalized 5$f$ energy level, unrenormalized conduction band, and hybridization strength, respectively. The representative fitting results are drawn in Fig.~\ref{fig:hyb}(c)-(d). Clearly, this model fits well at low temperatures. Then, the direct hybridization gap $\Delta$ is approximated by $2|V_k|$, while the indirect hybridization gap $\Delta'$ is evaluated by $|\epsilon_{+}(Z) - \epsilon_{-}(\Gamma)|$. Figure~\ref{fig:gap} displays $\Delta$ and $\Delta'$ as a function of $T$. Both gaps remain closed until $T = 290$~K, then abruptly open and increase gradually below 290~K. $\Delta'$ is quite small (only $1/20 \sim 1/15$ of $\Delta$) with a room temperature value of about 5~meV ($\approx 58$~K), in agreement with the previously measured value of $44 \pm 7$~K from pump-probe optical experiments~\cite{PhysRevLett.104.227002}.

Now let us concentrate on $P_{2}$ [see Fig.~\ref{fig:hyb}(e)-(h)]. The hybridization dynamics associated with this resonance differs from $P_{0}$. It exhibits the following traits: firstly, the $c-f$ hybridization is much weaker than $P_0$. At $T = 290$~K, only minor distortions affect the conduction bands $\zeta$ and $\eta$. Even at $T = 116$~K, the hybridization gap is not fully opened. 
Secondly, the simple mean-field model [i.e. Eq.~(\ref{eq:pam})] fails to accurately describe the hybridized bands within the considered temperature range. Thirdly, the hybridization strengths of $\zeta$ and $\eta$ bands are indeed different, with $V_{k,\zeta} \gg V_{k,\eta}$. This is evident that the $\zeta$ band is heavily distorted at $T = 290$~K, while the $\eta$ band remains almost linear. At $T = 116$~K, the $\zeta$ band opens a hybridization gap, whereas the $\eta$ band does not. This difference arises because the orbital character of the $\zeta$ band is mainly Ga$_1-s$, and that of the $\eta$ band is mainly Ga$_2-p_z$. As mentioned before, Ga$_1$ is at the center of $ab$ plane and is closer to the Pu atoms, making its $s$ orbitals more susceptible to $c-f$ hybridization. In contrast to $P_0$, the hybridization dynamics of $P_{2}$ exhibits an evident hybridization dynamics. Similar behaviors are observed for the other quasiparticle resonances ($P_1$, $P_3$, and $P_4$).

In summary, we employ a state-of-the-art first-principles many-body approach to investigate the correlated electronic structure of PuCoGa$_{5}$. A fingerprint of topology, specifically a Dirac node, is identified in the band structure. The coherent temperatures for the spin-orbit splitting of $5f_{5/2}$ states is found to be 580~K, suggesting a temperature-driven localized-itinerant crossover of 5$f$ states, which is higher than the opening of hybridization gaps at 290~K. The dynamics of quasiparticles and hybridization are influenced by strong valence state fluctuations. Our predictions reveal intriguing properties, including the unusual emergence and decay of high-energy quasiparticle resonances, distinct hybridization processes, and orbital-dependent Kondo screenings. Our findings shed new light on the the electronic structures, quasiparticle dynamics, and hybridization processes in $5f$ correlated electron systems. Further experimental and theoretical efforts are highly desired.

\begin{acknowledgments}
This work is supported by the National Natural Science Foundation of China (under Grant Nos.~12274380,~12474241, and~12434009), National Key Research and Development Program of China (under Grant No.~2022YFA1402201), and CAEP Project (under Grant No.~TCGH0710).
\end{acknowledgments}

\bibliography{pcgbib}

\end{document}